\documentclass[preprint,11pt]{emulateapj}  
\newcommand{\unit}[1]{\nobreak{\mathrm{\;#1}}} 
\newcommand{\mr}[1]{\nobreak{\mathrm{#1}}} 
\newcommand{\msc}[1]{\textrm{\textsc{#1}}}
\def\min{\textrm{\scriptsize{-}1}}
\def\dg{\textrm{\textit{\scriptsize{g}}}}
\newcommand{\ditto}[1]{{_{\rm{#1}}}} 
\newcommand{\ud}{{\rm d}} 

\newcommand{\be}{\begin{eqnarray}}
\newcommand{\ee}{\end{eqnarray}}
\newcommand{\eq}[1]{eq.~(\ref{eq:#1})}
\newcommand{\fig}[1]{Fig.~\ref{fig:#1}}
\newcommand{\uppo}[1]{{^{\rm{#1}}}} 

\begin{document}

\title{The Eddington Limit in Cosmic Rays: An Explanation for the 
Observed Lack of Low-Mass Radio-Loud Quasars and the $M_\bullet-M_\star$ Relation}
\author{Lorenzo Sironi and Aristotle Socrates\footnote{Current Address: Institute for Advanced Study, Princeton, NJ 08540}}
\affil{Department of Astrophysical Sciences, Princeton University, Princeton, NJ 08544-1001}
\email{lsironi@astro.princeton.edu; socrates@ias.edu}

\begin{abstract}
We present a feedback mechanism for supermassive black holes and  their host bulges
that operates during epochs of radio-loud quasar activity.  In the radio cores of relativistic quasar jets, internal shocks convert a fraction of ordered bulk kinetic energy into randomized relativistic ions, or in other words cosmic rays.  By employing a  phenomenologically-motivated jet model, we show that enough 1-100 GeV cosmic rays escape the radio core into the host galaxy to break the Eddington limit in cosmic rays. As a result, hydrostatic balance is lost and a cosmic ray momentum-driven wind develops, expelling gas from the host galaxy and thus self-limiting the black hole and bulge growth. Although the interstellar cosmic ray power is much smaller than the quasar photon luminosity, cosmic rays  provide a stronger feedback  than UV photons, since they exchange momentum with the galactic gas much more efficiently. The amount of energy released into the host galaxy as cosmic rays, per unit of black hole rest mass energy, is independent of black hole mass.  It follows that radio-loud jets should be more prevalent in relatively massive systems since they sit in galaxies with relatively deep gravitational potentials.  Therefore, jet-powered cosmic ray feedback not only self-regulates the black hole and bulge growth, but also provides an explanation for the lack of radio-loud activity in relatively small galaxies.  By employing basic known facts regarding the physical conditions in radio cores, we approximately reproduce both the slope and the normalization of the $M_{\bullet}-M_{\star}$ relation.
\end{abstract}


\section{Introduction}\label{sec:intro}
There are two classes of persistent sources at cosmic distances:
galaxies and quasars/Active Galactic Nuclei (AGN).    
Stars power galaxies while accretion onto and/or 
spin down of supermassive black holes power
quasars.  Until  recently, galactic and quasar
phenomena were thought to be separate on both observational 
and theoretical grounds.  However, the discovery of the black hole mass -- 
bulge stellar mass relation ($M_{\bullet}-M_{\star}$) in nearby elliptical galaxies 
\citep{magorrian_tremaine_98, mclure_dunlop_02, marconi_hunt_03, haring_rix_04} 
\begin{equation} 
M_{\bullet} \simeq 10^{-3} M_{\star}
\label{e: Magg}
\end{equation}
and the black hole mass -- stellar velocity dispersion relation
\citep[$M_{\bullet}-\sigma_{\star}$;][]{gebhardt_00, ferrarese_merritt_00,tremaine_02}
\begin{equation}\label{e: Magg2}
M_{\bullet}\propto \sigma^4_{\star}
\end{equation}
indicates that galactic and black hole activity are closely connected
to one another. The natural implication is that the energy release resulting 
from the build-up of the black hole mass 
limits any further growth of both the stellar bulge
and the black hole.  The fact that the
$M_\bullet-M_{\star}$ relation holds for nearly four decades in black hole mass seems to suggest that a {\it universal}, {\it self-similar} 
or {\it scale-free} process is at work, which acts to self-regulate
the ratio between black hole and bulge mass, irrespective
of their combined mass.  Apparently, the only 
question that remains is with regard to the
exact physical mechanism responsible for 
black hole feedback and self-regulation.    

The \citet{soltan_82} argument, along with the 
work of \citet{yu_tremaine_02}, indicates that the 
 mass of supermassive black holes is mostly accrued during an optically-luminous radiatively-efficient ``quasar phase.'' The energy released during the accretion process, which is carried away primarily by photons and/or a ``quasar wind,''  may couple to the  interstellar medium of the host galaxy and eject it from the galactic gravitational potential \citep[e.g.,][]{silk_rees_98, fabian_99, ciotti_ostriker_01, king_03,dimatteo_05,murray_05, hopkins_06}. 
In doing so, fuel for any 
further galactic and quasar activity is removed and the mass of 
the black hole, as well as of the stellar bulge, is self-limited. 

The energy released during the accretion process may be carried away not solely in the form of photons.  In the so called ``radio-loud'' (as opposed to ``radio-quiet'') objects, relativistic collimated
outflows, or ``jets,'' put out a significant
amount of energy in mechanical form. Although radio-loud phenomena are also observed in objects that are not actively accreting \citep[e.g.,][]{ho_peng_01,ho_02}, the radio jet is more likely to affect the evolution of the system when a significant amount of mass is being built up. As  this work focuses on the self-regulation of black hole growth, which occurs at high accretion rates, our attention rests on objects that are both significantly accreting and radio-loud. 

The kinetic power of radio jets is dissipated in sub-pc scale ``radio cores'' and kpc to Mpc scale 
``radio lobes,'' with comparable amounts of energy dissipated at each
site. The radio-loud quasar phase could be responsible for black hole 
self-regulation if the energy release from the radio core, 
unlike that from the  distant radio lobes, has the opportunity to 
 couple to the interstellar medium of the host galaxy.

In this work, we propose that black hole self-regulation may be mediated by 1-100 GeV protons, or in other words \textit{cosmic rays}, produced in the jet core during phases of radio-loud quasar activity. We show that enough cosmic rays escape the radio core into the host galaxy to power a cosmic ray-driven wind that ejects the interstellar gas, thus removing the fuel for  further star formation and black hole accretion. 

Our arguments are organized as follows.
In \S\ref{s: basic} we lay out the basic idea behind our jet-powered cosmic
ray feedback scenario by defining and recognizing the importance
of the Eddington limit in cosmic rays.  In
\S\ref{s: model}  we briefly summarize the physical features of the standard model of
radio-loud AGN jets that are important for our feedback mechanism.  With the jet model in hand,  in \S\ref{s: eff_lum} we determine the
interstellar cosmic ray luminosity resulting from radio-loud quasar
activity in terms of observable quantities. For readers uninterested in the details of our jet and  radio core model, as well as its phenomenological underpinnings, we suggest skipping \S\ref{s: model} and \S\ref{s: eff_lum}. 
In \S\ref{sec:feedback} we examine the
consequences of our cosmic ray feedback scenario and make a critical
comparison of the black hole self-regulation model presented here
with other models of quasar feedback. We summarize our findings in \S\ref{s: test}, where we also present a brief review of well-known
observations that lend support to our jet-powered
self-regulation mechanism. In addition, we
propose an observational test. 

\section{The Eddington Limit in Cosmic Rays}
\label{s: basic}
If supermassive black hole growth results primarily from 
radiatively-efficient accretion, as the \citet{soltan_82} argument suggests, then the ratio of the 
total energy released during the accretion process 
 to the binding energy of the 
galaxy's gaseous phase is given by 
\begin{eqnarray}
\!\!\!\!\!\!\!\!\frac{\Delta E_{\bullet}}{E_g} &\simeq& \frac{\epsilon_\ditto{rad}\,M_{\bullet}c^2}{
f_gM_{\star}\sigma^2_{\star}}\simeq10^3\!\left(\frac{M_{\bullet}/M_{\star}}{10^{-3}}\right)\!\epsilon_\ditto{rad,\min}\,f_{g,\min}^{-1}\,\sigma_{\star,300}^{-2}~,
\label{e: DeltaE}
\end{eqnarray}
in case the black hole -- galaxy system follows the $M_\bullet-M_\star$ relation in eq.~(\ref{e: Magg}). Here, $\epsilon_\ditto{rad}=0.1\,\epsilon_\ditto{rad,\min}$ is the radiative efficiency of the accretion flow. The binding energy of the gas $E_g\simeq f_gM_{\star} \sigma^2_{\star}$ is appropriate for a galaxy resembling an isothermal sphere whose gravitational force is mainly provided by stars and dark matter; $f_g=0.1\,f_{g,\min}$ is the galaxy's gas fraction and $\sigma_\star=300\,\sigma_{\star,300} \unit{km\,s^{-1}}$ is the stellar velocity dispersion. Clearly, only a small amount of the photon energy  released during optically-bright quasar epochs is needed to couple to 
the galactic gas in order to eject it from the galaxy's
 gravitational potential, thus self-limiting the combined growth of  the black hole and the stellar bulge. 

Again, during radio-loud phases a significant amount of the energy resulting from black hole accretion is released in mechanical form as powerful relativistic jets.  Even if the time-integrated kinetic output of the jet is $\Delta E_\ditto{J} \ll 
\Delta E_{\bullet}$, eq.~(\ref{e: DeltaE}) indicates that
there may be ample energy to 
unbind the galactic gas, provided that the fraction of jet kinetic energy deposited in the radio core can efficiently couple to the gaseous component of the host galaxy. In other words, radio-loud kinetically-dominated epochs may represent another mode of AGN activity, other than optically-luminous radiatively-efficient quasar phases, that could potentially lead to efficient black hole self-regulation.

The photon luminosity of the sub-pc scale jet core results from the cooling of relativistic
electrons through a combination of synchrotron and inverse Compton
emission. By modeling the spectral energy distribution of powerful radio-loud quasars, \citet[][]{celotti_ghisellini_07} infer that the emitting
electrons must be accelerated, presumably by some magnetized
Fermi mechanism, up to highly-relativistic energies, in excess of a few tens of GeV (in the jet comoving frame). 
The details of the acceleration mechanism aside, it is reasonable to conclude  as well that  protons are randomized with energies up to $\sim1-10\unit{GeV}$. If a significant fraction
of these randomized relativistic ions,
or in other words {\it cosmic rays}, escape the site of
acceleration and diffuse into the interstellar medium of the host galaxy, they may provide the coupling between jet power and galactic gas required for black hole self-regulation during radio-loud phases.

For example, in actively star-forming galaxies the generation and subsequent diffusion of cosmic ray protons -- 
a well-known by-product of core-collapse supernovae, 
and thus massive star-formation --
may act as an agent of self-regulation, limiting the rate of 
star-formation 
and therefore the galaxy luminosity. 
The simplest way to understand this is by defining the 
{\it Eddington limit in cosmic rays} 
\begin{eqnarray}
 \!\!\!\!\!L_{_{\rm Edd, CR}} &=& 
L_{_{\rm Edd, \star}}\frac{\lambda_\ditto{CR}}{
\lambda_\ditto{T}}\nonumber\\
&\simeq &1.3\times10^{44}\!\left(
\frac{\lambda_\ditto{CR}/\lambda_\ditto{T}}{10^{-6}}\right)\!\!
\left(\frac{M_{\star}}{10^{12}M_{\odot}}\right)\!\unit{erg\,s^{-1}}\,,
\label{e: LeddCR}
\end{eqnarray}
where $L_{_{\rm Edd, \star}}$ is the Thomson Eddington limit for a galaxy with mass $M_\star$; we have pinned the ratio between the cosmic ray mean free path and the
Thomson mean free path  to the value  $\lambda_\ditto{CR}/\lambda_\ditto{T} \sim
10^{-6}$, as in the case of the Milky Way.
An interstellar cosmic ray luminosity of $L_\ditto{CR}\sim 10^{44}\,{\rm
erg\,s^{-1}}$ -- equal to the corresponding cosmic ray Eddington limit $L_{_{\rm Edd,
CR}}$ for the most massive galaxies -- may result from the act of
forming stars at a rate of $\sim 10^3\,{M}_{\odot}\,{\rm yr^{-1}}$
-- corresponding to the brightest star-forming galaxies.  The formation 
of stars at a higher rate leads to the breaking of the 
cosmic ray Eddington limit and a cosmic ray-driven wind develops, removing the gaseous phase 
of the galaxy thus
choking off star-formation.  From this straightforward
argument, it is quite possible that the luminosities of 
star-forming galaxies are capped at their observed values
because they are Eddington-limited in cosmic rays, as proposed by \citet{socrates_06}.

With respect to AGN, radio-loud objects display powerful jets with kinetic luminosity upwards of
$L_\ditto{J}\sim 10^{47}\, {\rm erg\,s^{-1}}$  \citep[e.g.,][]{rawlings_saunders_91}, comparable to the Thomson Eddington limit $L_\ditto{Edd,\bullet}$ for the most massive black holes with
$M_{\bullet} \sim 10^9\, M_{\odot}$.  If the jet kinetic
power is a fraction $\Lambda_{_{
\rm Edd}}\sim1$ of the black hole Thomson Eddington limit, we have 
\begin{eqnarray}
L_\ditto{J}=\Lambda_{_{\rm Edd}}L_{_{\rm Edd,\bullet}}\simeq1.3\times10^{47}\Lambda_{_{\rm Edd}}
M_{\bullet,9} \unit{erg\,s^{-1}}~,
\label{e: Ljet}
\end{eqnarray}
where
$M_{\bullet}=10^9M_{\bullet,9}\,M_{\sun}$. Let $\epsilon_\ditto{CR}$ be the efficiency for converting the jet kinetic power $L_\ditto{J}$ into interstellar cosmic ray  
power $L_\ditto{CR}$. A statement of momentum
conservation and hydrostatic balance, the Eddington limit in cosmic
rays defined in eq.~(\ref{e: LeddCR}) indicates that a galactic cosmic ray-driven wind develops if
\begin{equation}\label{eq:balance}
L_\ditto{CR}= \epsilon_\ditto{CR}\,L_\ditto{J} \gtrsim L_{_{\rm Edd, CR}}~,
\end{equation}
or if the fraction of jet kinetic luminosity ending up as interstellar cosmic rays attains a value of
\begin{equation}
\epsilon_\ditto{CR}\gtrsim 
10^{-3}\,\Lambda_{_{\rm Edd}}^{-1}
\left(\frac{\lambda_\ditto{CR}/\lambda_\ditto{T}}{10^{-6}}\right)
\left(\frac{M_{\star}/M_{\bullet}}{10^3} \right)
\label{e: min_eff}
\end{equation}
for a black hole -- galaxy system that follows the $M_\bullet-M_\star$ relation.

The \textit{momentum} requirement in eqs.~(\ref{eq:balance}) or (\ref{e: min_eff}) is a necessary condition to initiate a galactic cosmic ray-driven outflow. However, in order to fully unbind the galaxy's gaseous phase, thus self-regulating the black hole and bulge growth, this epoch of super-Eddington cosmic ray activity should last long enough so that the time-integrated \textit{energy} injected as cosmic rays into the interstellar medium is comparable to the binding energy of the galactic gas. Interestingly, the value of the jet-cosmic ray conversion efficiency
 given  in eq.~(\ref{e: min_eff}) is
roughly equal to $E_g/\Delta E_{\bullet}$ for the most massive galaxies (see eq.~(\ref{e: DeltaE})), where $E_g/\Delta
E_{\bullet}$ may be viewed as the minimum efficiency of a ``generic'' feedback mechanism for
coupling the black hole energy release to the gaseous
component of the host galaxy.  In other words,  the
minimum value of $\epsilon_\ditto{CR}$ required to break the galaxy's {\it momentum}
balance, given in eq.~(\ref{e: min_eff}), is approximately equal to 
the minimum value of the {\it energy} coupling efficiency 
$E_g/\Delta E_{\bullet}$ for the most massive systems.\footnote{Strictly speaking, the minimum energy coupling efficiency for our jet-powered cosmic ray feedback would be $\sim E_g/\Delta E_\ditto{J}$, where $\Delta E_\ditto{J}$ is the time-integrated kinetic output of the radio jet. We are implicitly assuming that  $\Delta E_\ditto{J}\sim\Delta E_\bullet$ at the high-mass end.} 

Clearly, an understanding and determination of the jet-cosmic ray
efficiency parameter $\epsilon_\ditto{CR}$ is of fundamental 
importance with respect to quantifying whether or not  
jet-cosmic ray feedback is in fact the agent of self-regulation 
for a given black hole -- galaxy system. In what follows,
we take into account the physical properties of AGN 
radio cores with the help of a well-established
jet model.  In doing so, we are able to determine 
the jet-cosmic ray efficiency parameter $\epsilon_\ditto{CR}$ and discuss the effectiveness of our cosmic ray-driven feedback scenario.

\section{A Standard Jet Model for Radio-Loud AGN}
\label{s: model}
Here we describe the physical properties of quasar jets, whose radio core is responsible for injecting cosmic rays into the interstellar medium of the host galaxy. The reasoning and details of the jet model outlined below are grounded by decades of observations of radio-loud quasars. We also discuss some of the observational constraints that we use to determine the parameters of our model.

\subsection{Basics of Radio-Loud AGN Phenomena}
The spectral signature of
radio-loud AGN results from dissipation of the kinetic energy of powerful relativistic
jets in sub-pc scale ``radio cores'' and kpc to Mpc scale ``radio lobes.''
The basic radio-loud phenomenology is largely understood in
terms of a viewing-angle effect \citep[e.g.,][]{antonucci_93, urry_padovani_95}.  Emission from the relativistic jet flow in the radio core is highly beamed
due to relativistic aberration. The radiation seen along
the jet axis is Doppler-boosted in both frequency and flux, when compared 
to the flow rest frame, and the opposite is true when the radio core
is viewed edge-on. 

Core-dominated on-axis sources are collectively referred 
to as ``blazars'' -- a marriage between BL Lacs
and flat-spectrum radio quasars (FSRQs).  They display extreme variability, 
broad (radio to gamma-ray) spectral 
energy distributions, super-luminal motion 
and super-Eddington fluxes. BL Lac spectra are typically featureless, whereas intense line emission is detectable in the optical-UV spectrum of FSRQs, presumably arising from the accretion disk and its associated broad-line region (BLR). The continuum spectral energy distribution of blazars shows two prominent bumps, with the high-energy bump that often dominates the bolometric power output. The low-energy bump, at infrared to UV frequencies, is conventionally attributed to synchrotron radiation, while the high-energy emission, extending up to gamma-ray energies, is thought to result from inverse Compton up-scattering of synchrotron photons (in BL Lacs) or external photons from the accretion disk and the BLR (in FSRQs).

In
objects where the jet is viewed edge-on, the so called ``Fanaroff-Riley'' galaxies, the radio emission mostly results from the dissipation of jet power in radio lobes on inter-galactic or super-galactic
scales. The emission from sub-pc scales is largely beamed away from the observer for 
these off-axis sources. In the framework of unification models  \citep[e.g.,][]{antonucci_93, urry_padovani_95}, FR I radio galaxies are identified as the parent population of BL Lacs, while FR II galaxies are the edge-on counterpart of FSRQs.

In this work, all of our attention will be directed towards the most powerful objects, i.e., FSRQs and FR II galaxies, sources where radio-loud activity coexists with substantial black hole accretion. From here on, FSRQs and their parent population of FR II radio galaxies will be collectively referred to as ``radio-loud quasars''.

\subsection{A Simple Jet Model}
\label{sec:jetmodel}

We assume that the jet power or ``luminosity'' $L_\ditto{J}$ is primarily 
carried in mechanical form by matter expelled from the central engine at
a rate $\dot{M}_\ditto{J}$, such that
\begin{equation}
L_\ditto{J}\simeq\Gamma\dot{M}_\ditto{J}c^2~,
\end{equation} 
where $\Gamma$ is the characteristic Lorentz factor of the flow.  Under
the assumption that the ultimate 
source of jet power is accretion onto 
the black hole at a rate $\dot{M}_{\bullet}$ (as opposed to extraction of the black hole spin), then 
$\dot{M}_\ditto{J}/\dot{M}_{\bullet}\sim \epsilon_\ditto{rad}/\Gamma$ if the jet kinetic power is comparable to the accretion photon luminosity,
$L_\ditto{J}\sim L_\ditto{acc}\simeq\epsilon_\ditto{rad}\dot{M}_{\bullet}c^2$.  It follows that for 
typical values, namely $\epsilon_\ditto{rad}\sim 0.1$  and
$\Gamma\sim 10$ \citep[e.g.,][]{jorstad_05}, we have $\dot{M}_\ditto{J}/\dot{M}_{\bullet}\sim
0.01$.  In other words, a significant fraction of the accretion 
power must be transmitted to only $\sim 1\%$ of the mass.  
The mechanism of this seemingly improbable  transfer of energy is assumed to be an unknown from here on as we 
are primarily interested in its mechanical output.   

Powerful radio-loud quasars are variable, often 
displaying radio blobs, or ``knots,'' moving along the jet.  Such behavior 
is commonly thought to result from the dissipative
interaction of shells of matter intermittently ejected from the central engine. If the 
Lorentz factors of the shells differ such that $\Delta\Gamma\sim\Gamma$ in the reference frame of the host galaxy (here, $\Gamma$ may be thought of as the mean Lorentz
factor of the jet flow), then the characteristic 
distance $R_\ditto{diss}$ from the black hole where dissipation takes place is \begin{equation}\label{eq:r_diss}
R_\ditto{diss}\simeq \frac{\Gamma^2}{\epsilon_\ditto{rad}}R_\ditto{G}\simeq
0.05\,\,\Gamma^2_1\,\epsilon_\ditto{rad,\min}^{-1} M_{\bullet,9}\unit{pc}~,
\end{equation}            
where $\Gamma=10\,\Gamma_1$; $R_\ditto{G}=G M_\bullet/c^2$ is the black hole gravitational radius and
$R_\ditto{G}/\epsilon_\ditto{rad}c$ is the dynamical timescale of the accretion flow, which determines the typical time delay between subsequent
shell ejections by the central engine. For the sake of simplicity, we adopt a one-zone model in which the whole jet spectral energy distribution originates from a single region located at $R_\ditto{diss}$.

We assume that the jet is beamed into a cone with half-opening
angle $\theta\sim0.1$ \citep[e.g.,][]{jorstad_05}. For typical values $\Gamma\sim10$ and $\theta\sim0.1$, in the jet comoving frame the dissipation region resembles  a spherical blob, since its comoving longitudinal length $\sim R_{\ditto{diss}}/\Gamma$ is comparable to its transverse size $\sim R_\ditto{diss}\theta$. For a bi-conical mass-dominated jet,
\begin{equation}\label{eq:lljet}
L_\ditto{J}\simeq\Gamma\,\dot{M}_\ditto{J}c^2\simeq2\pi(R\,\theta\;\!)^2\Gamma\rho\,c^3~,
\end{equation}
where $R$ is the distance from the black hole and $\rho$ is the jet mass density at radius $R$ measured in the host-galaxy frame. If the jet kinetic power is a fraction $\Lambda_{_{\rm Edd}}$ of the black hole's photon Eddington limit, the density $\rho_{_{\rm diss}}$ at the dissipation scale $R_\ditto{diss}$
and the Thomson optical depth $\tau_{_{\rm diss}}$ down to $R_\ditto{diss}$  may be written in terms of basic physical parameters as
\begin{equation}\label{eq:rho_diss}
\rho_{_{\rm diss}}\simeq \frac{2\,\epsilon_\ditto{rad}^2\Lambda_{_{\rm Edd}}}{\Gamma^5
\theta^2\kappa_{es}R_\ditto{G}}
\end{equation}
and 
\begin{eqnarray}\label{eq:tau_diss}
\!\!\!\!\!\!\!\tau_{_{\rm diss}}\simeq \frac{2\,\epsilon_\ditto{rad}\Lambda_{_{\rm Edd}}}{
\Gamma^3\,\theta^2}\frac{n_e}{n_p}\simeq0.02\,\frac{n_e}{n_p}\,\epsilon_\ditto{rad,\min}\Lambda_{_{\rm Edd}}\Gamma_1^{-3}\theta^{-2}_{\min}~,
\end{eqnarray}
where $\theta=0.1\,\theta_{\min}$ and $\kappa_{es}\simeq0.4\unit{cm^2\,g^{-1}}$ is the electron scattering opacity; $n_p$ and $n_e$ are the jet number densities of protons and leptons (including both positrons and electrons) in the host-galaxy frame. In \eq{tau_diss} we assume that the jet kinetic power is mostly carried by protons -- in \S\ref{sec:content} we estimate that pairs outnumber ions by only a factor of ten. For $n_e/n_p\sim10$, it follows from \eq{tau_diss} that the dissipation region is optically thin.

In the dissipation region, internal shocks resulting from shell-shell collisions convert a fraction $\epsilon_\ditto{th}$ of the orderly mechanical energy of the jet flow into heat. It follows that the randomized thermal power generated in the dissipation region is
\begin{equation}
L_\ditto{th}= \epsilon_\ditto{th}L_{_{\rm J}}~,
\end{equation}
which corresponds to a thermal energy density 
in the comoving frame
\begin{equation}\label{eq:eps_p}
U_\ditto{th}'=\epsilon_\ditto{th}\,\rho_{_{\rm diss}}'\,c^2
=\frac{\epsilon_\ditto{th}}{\Gamma}\rho_{_{\rm diss}}c^2~,
\label{e: U_cr}
\end{equation}  
where $\rho'_\ditto{diss}=\rho_{_{\rm diss}}/\Gamma$ is the jet mass density at $R_\ditto{diss}$ measured in the comoving frame. If the two colliding shells have equal rest mass and differ by $\Delta\Gamma$ in Lorentz factor, the fraction of bulk kinetic energy converted into heat is $\epsilon_\ditto{th}\simeq1/8\,(\Delta\Gamma/\Gamma)^2$ to leading order in $\Delta\Gamma/\Gamma$ \citep[e.g.,][]{kobayashi_97}. We take $\epsilon_\ditto{th}\sim0.1$ as a benchmark for shells with $\Delta\Gamma\sim\Gamma$.  It follows that the average comoving Lorentz factor of shocked protons is $\gamma_p\simeq1+\epsilon_\ditto{th}$, corresponding to a mean energy $\simeq\Gamma\,\gamma_p\,m_pc^2\sim10\unit{GeV}$ in the frame of the host galaxy. If a sufficient number of these randomized relativistic ions, or cosmic rays,  escape the jet into the galaxy's interstellar medium, they may provide the coupling between jet power and galactic gas required for our feedback model.

Since radio cores are inarguably 
powered by synchrotron emission, our jet model cannot be complete
without an estimate of the magnetic field.  We parametrize the magnetic 
energy density in the comoving 
frame $U_{_{\rm B}}'$ as a fraction $\epsilon_{_{\rm B}}\sim0.1$ of the thermal energy density $U_\ditto{th}'$:
\begin{equation}
U_{_{\rm B}}'=\epsilon_{_{\rm B}}U_\ditto{th}'~.
\label{e: eps_B}
\end{equation}
It is reasonable to suspect that the magnetic field is 
 tangled and inhomogenous within the dissipation region, more so since we assume the jet electromagnetic flux  to be sub-dominant with respect to the kinetic flux. In the comoving frame, the mean magnetic field strength is 
\begin{equation}\label{eq:Bfield}
B'\simeq 2.7 \;\epsilon^{1/2}_{\ditto{B,\min}}\,\epsilon^{1/2}_{\ditto{th,\min}}\,\epsilon_{\ditto{rad,\min}}
\Lambda^{1/2}_{_{\rm Edd}}\,\Gamma^{-3}_1\,
\theta^{-1}_\min M^{-1/2}_{\bullet,9}\,{\rm G}~,
\end{equation}
where $\epsilon_\ditto{th}=0.1\,\epsilon_\ditto{th,\min}$  and $\epsilon_\ditto{B}=0.1\,\epsilon_\ditto{B,\min}$. 

Now, we possess all of the necessary physical 
ingredients for our feedback model.  However, before 
we assess whether or not a sufficient number of cosmic ray protons 
are able to leak out of the radio core, potentially leading 
to the disruption of the entire host galaxy, we spend some time
 -- for the sake of completeness -- on the observational 
motivations that went into our jet model.  

\subsection{Observational Constraints on the Jet Model} \label{sec:obs_model}

Many of the basic physical features of the somewhat ``canonical''
jet model outlined above are strongly supported and motivated by 
decades of observations.  A synopsis of the corroborating 
evidence follows.    

\subsubsection{Jet Kinetic Luminosity:  $L_{_{\rm J}}$
and $\Lambda_{_{\rm Edd}}$}

Edge-on Fanaroff-Riley jets are better suited than head-on blazars to measure the jet kinetic power. For blazars, the highly Doppler-beamed, variable, broad-band radiation output must be properly modeled in order to extract physical parameters at 
the jet dissipation scale.  At any given frequency, the value of the comoving 
radiation flux $F'_{\nu'}$ itself is difficult to determine.  In fact, 
in the observer frame, 
\begin{equation}
F_{\nu}\simeq \delta^3 F'_{\nu'}~,
\end{equation} 
where $\delta$ is the Doppler-beaming factor 
\citep[e.g.,][]{lind_blandford_85}, which depends upon 
$\Gamma$ and the orientation of the observer's line of 
sight with respect to the jet axis. Clearly, due to the 
strong dependence on the Doppler-beaming factor $\delta$, 
an accurate measurement of the comoving radiation 
flux $F'_{\nu'}$, and therefore
of intrinsic jet properties such as $L_{_{\rm J}}$
and $\Lambda_{_{\rm Edd}}=L_\ditto{J}/L_\ditto{Edd,\bullet}$,
is difficult to perform.  

In the case of Fanaroff-Riley sources, the terminal 
sites of jet dissipation, the so called radio lobes,
can be thought of as the thermal reservoirs of 
the initially ordered and collimated jet mechanical energy.  Since the
energy injection rate outstrips the radiative cooling 
rate, radio lobes have no choice other than to expand.  
Plasma temperature and density within a radio lobe
are typically extrapolated from its X-ray continuum and 
line emission. From the inferred lobe enthalpy $E_\ditto{lobe}$, the jet kinetic power can be estimated as  
\begin{equation}
L_{_{\rm J}}\simeq \frac{E_{_{\rm lobe}}}{t_{_{\rm lobe}}}
 \simeq\frac{E_{_{\rm lobe}}\,c_s }{l_{_{\rm lobe}}}~,
\end{equation}
where $t_{_{\rm lobe}}\simeq l_{_{\rm lobe}}/c_s$ is the characteristic time required to inflate a lobe with width $l_{_{\rm lobe}}$ at the sound speed $c_s$ \citep[e.g.,][]{birzan_04, shurkin_07}.

At the powerful end, the lobes of FR II galaxies possess average kinetic inputs of 
$L_{_{\rm J}}\simeq 10^{47}-10^{48}\,{\rm erg\,s^{-1}}$, corresponding
to the Thomson Eddington limit for  black holes with mass $10^9-10^{10}\,M_{\odot}$.\footnote{Black-hole mass is usually estimated from optical and UV lines, since their width supposedly measures the depth of the
gravitational potential at the BLR.}
The fact that the upper limit for $L_{_{\rm J}}$ is close to the
Eddington limit for the largest black holes informs us that, at least
during periods of powerful activity, a value of $\Lambda_{_{\rm Edd}}\simeq 
1$ may be appropriate for these sources, and that $\Lambda_{_{\rm Edd}}\sim 1$ may serve
as a reasonable upper bound. For powerful FSRQs, thought to be the on-axis counterpart of massive FR II galaxies, \citet{celotti_ghisellini_07} infer comparable values for $L_\ditto{J}$  from modeling the blazar spectral energy distribution.

\subsubsection{Optical Depth and Pair Content: $\tau_{_{\rm diss}}$
and $n_e/n_p$}\label{sec:content}

Information on particle composition at the jet dissipation scale is best extracted from the spectra of FSRQs, where both jet and disk emission
are present. The general procedure is outlined by \citet{sikora_97}.  

The jet is bathed by optical/UV photons from the
BLR and IR photons from the presumed obscuring
torus. Owing to the large bulk Lorentz factor $\Gamma\sim10$ of the jet flow, ``cold'' electrons in the jet have the capacity
to Compton up-scatter these relatively soft photons (a process known as ``bulk-Comptonization'' or first-order kinetic Sunyaev - Zel'dovich effect) up to characteristic energies
\begin{equation}\label{eq:bc_freq}
 h\nu_{_{\rm BC}}\simeq \Gamma^2h\nu_{_{\rm UV}}
\simeq \Gamma^2_1\;{\rm keV}~,
\end{equation}
where
$h\nu_{_{\rm UV}}\simeq10\unit{eV}$ is the typical seed photon energy. The expected bulk-Compton luminosity $L_\ditto{BC}$ from cold jet electrons may be written as a volume integral extending from the jet base $\sim R_\ditto{G}/\epsilon_\ditto{rad}$ to the dissipation scale $R_\ditto{diss}$, i.e., below the region where electrons are shock-heated:
\begin{eqnarray}
L_{_{\rm BC}}&\simeq& \frac{\delta^3}{\Gamma}
\int\ud Vn_e\left|\frac{\ud E_e}{\ud t}\right|\simeq 
2\Gamma^2\int\, \ud Vn_e\,c\,\sigma_\ditto{T}\Gamma^2\,U_\ditto{BLR}\nonumber\\
&=&\frac{1}{2}\Gamma^4\theta^2\, \xi\,L_{_{\rm acc}}\tau_{_{\rm cold}}~,
\label{e: bulk_Comp}
\end{eqnarray} 
where $\ud E_e/\ud t$ is the rate at which a cold electron 
loses its orderly kinetic energy via the bulk-Comptonization process, $\sigma_\ditto{T}$ is the Thomson cross section and $U_\ditto{BLR}=\xi\,L_{_{\rm acc}}/(4\pi R^2c)$ is the energy density at radius $R$ resulting from the fraction $\xi$ of accretion luminosity $L_\ditto{acc}$ re-processed and isotropized by the BLR.
Eq.~(\ref{e: bulk_Comp})
leads to an expression for the Thomson optical 
depth $\tau_\ditto{cold}$ from the jet base up to the dissipation region:
\begin{eqnarray}\label{eq:bc_lum}
\!\!\!\!\tau_{_{\rm cold}}  \simeq  \frac{2}{\Gamma^4\,\theta^2}  \frac{L_{_{\rm BC}}}
{\xi\,L_{_{\rm acc}}}
\lesssim 0.2\left(\frac{L_{_{\rm SX}}
/\xi L_{_{\rm acc}}}{10}\right)\Gamma_{1}^{-4}
\theta^{-2}_{\min}~,
\end{eqnarray}
where we have imposed that the expected bulk-Compton luminosity $L_\ditto{BC}$ should not exceed the observed integrated power $L_{_{\rm SX}}$ at soft X-ray energies (where the bulk-Compton emission should peak, see \eq{bc_freq}).

Assuming that the flux of cold electrons is conserved
along the jet (i.e., pair injection occurs primarily at the jet base), it follows that the Thomson depth $\tau_\ditto{diss}$ down to the dissipation scale satisfies $\tau_{_{\rm diss}}\lesssim\tau_{_{\rm cold}}$, since the optical depth is greater at smaller radii. Thus, from eqs.~(\ref{eq:tau_diss}) and (\ref{eq:bc_lum}) the electron to proton ratio of the jet flow is constrained to be
\begin{eqnarray}
\frac{n_e}{n_p}\lesssim 10 \left(\frac{L_{_{\rm SX}}/\xi 
L_{_{\rm acc}}}{10}\right)\Lambda_{_{\rm Edd}}^{-1}\epsilon_{_{\rm rad,\min}}^{-1}\Gamma_1^{-1}~,
\label{e: pair_frac}
\end{eqnarray}
which informs us that electrons and positrons, though 
dominant by number, only
advect with them an insignificant fraction of the jet power.
That is, the kinetic power of the jet is almost  
entirely carried by ions, as assumed in \S\ref{sec:jetmodel}. Also, from \eq{tau_diss} with $n_e/n_p\sim10$ it follows that $\tau_\ditto{diss}\simeq0.2$, and the flow in the dissipation region is optically thin.

\subsubsection{Magnetic Field Strength: $\epsilon_{_{\rm B}}$}

As in the case of estimating the jet composition,
\mbox{FSRQs} are the best sources for measuring the relative
strength of the magnetic field at the 
dissipation scale, parametrized by $\epsilon_{_{\rm B}}=
U'_{_{\rm B}}/U'_\ditto{th}$.  In the event 
that all of the low-energy synchrotron (with integrated luminosity $L_\ditto{S}$) and high-energy inverse Compton ($L_\ditto{IC}$) 
emission is powered by electrons accelerated in 
the dissipation region, then 
\begin{equation}
\frac{L_{_{\rm S}}}{L_{_{\rm IC}}}\simeq
\frac{U'_{_{\rm B}}}{U'_{_{\rm soft}}}=
\epsilon_{_{\rm B}}
\frac{U'_\ditto{th}}{U'_{\ditto{soft}}}~,
\end{equation}
where $U'_{\ditto{soft}}$ is the comoving energy density of seed photons for the inverse Compton process. For FSRQs, $U'_{\ditto{soft}}$ is mostly contributed by external photons from the BLR and obscuring torus (as opposed to synchrotron seed photons, that dominate in BL Lacs), so that $U'_\ditto{soft}\sim\Gamma^2\,U_\ditto{BLR}$, where $U_\ditto{BLR}$ was defined in \S\ref{sec:content}. With 
help from 
eq.~(\ref{e: U_cr}) for the form of $U'_\ditto{th}$
we have
\begin{equation}
\frac{U'_\ditto{th}}{U'_{\ditto{soft}}}
\simeq \frac{2\,\epsilon_\ditto{th}}{\Gamma^4\theta^2\xi}~,
\end{equation}
where we made liberal use of the fact that
$ L_{_{\rm J}}\sim L_{_{\rm acc}}$ for our jet model in 
FSRQs.  It follows that
\begin{equation}\label{eq:mag_frac}
\epsilon_{_{\rm B}}\simeq 0.05\left(\frac{L_{_{\rm S}}
/L_{_{\rm IC}}}{0.1}\right)
\left(\frac{\xi}{10^{-3}}\right)\epsilon_\ditto{th,\min}^{-1}\Gamma_1^4\,\theta_{\min}^2~,
\end{equation}     
which is consistent with the value $\epsilon_\ditto{B}\sim0.1$ chosen in \S \ref{sec:jetmodel}. Comparable values result from detailed modeling of blazar spectral energy distributions \citep{celotti_ghisellini_07}. This confirms that the electromagnetic contribution to the jet energy flux at the dissipation scale is negligible compared to the proton kinetic flux.

The magnetic energy fraction $\epsilon_\ditto{B}$ computed in \eq{mag_frac} is independent from the black hole mass  
$M_{\bullet}$.  The apparent lack of dependence of $\epsilon_{_{\rm B}}$
on $M_{\bullet}$ may be misleading.  In fact, the fraction $\xi$ of 
accretion luminosity that is re-processed and isotropized in the
BLR will in principle depend upon $M_{\bullet}$, since 
the characteristic disk temperature scales as $ 
\propto M^{-1/4}_{\bullet}$
at fixed Eddington ratio \citep{shakura_sunyaev_73}, which 
leads to a corresponding change in the photo-ionizing luminosity
per unit of accretion power. Nevertheless, for the level of 
accuracy of this work we ignore such complications.    

\subsection{Summary of the Jet Model}
Our ``canonical'' jet model may be considered more
as a ``consensus'' jet model.  That is, it is based upon decades of observations and modeling 
of radio-loud quasars, rather than any deep theoretical principle.  We do not address the nature of the 
ultimate source of mechanical energy at the base of the jet.
We simply exploit the fact that the jet is launched with some 
mixture of protons and electrons at a constant Lorentz factor 
$\Gamma\sim 10$ and half-opening angle $\theta\sim 0.1$, as 
 often inferred in bright core-dominated
sources. Although pairs may outnumber protons, the jet power is dominated by the ions. The most important feature of the jet model 
in terms of cosmic ray production is that dissipation by 
internal shocks at a distance $R_{_{\rm diss}}\simeq \Gamma^2
R_\ditto{G}/\epsilon_\ditto{rad}$ from the black hole 
leads -- almost certainly on theoretical 
grounds -- to the randomization of bulk baryonic 
energy and -- without doubt on observational grounds --
of bulk leptonic energy. The shock-heated protons that escape the jet into the interstellar medium of the host galaxy may have profound implications for the galaxy evolution.

The most important aspect of our jet model 
is that it is extremely simple with relatively 
few inputs.  That is, we adopt typical values for
$\Gamma$, $\Delta\Gamma$, $\theta$  and $\epsilon_{_{\rm rad}}$ and we derive physically-motivated estimates for $
\Lambda_{_{\rm Edd}}$, $\epsilon_\ditto{th}$ and  $\epsilon_\ditto{B}$.
From these parameters, we 
extract all the necessary information required to 
model our jet-powered cosmic ray feedback mechanism, as we describe in the next section.
   
\vspace{-0.03in}
\section{Determination of $\epsilon_{_{\rm CR}}$ and $L_{_{\rm CR}}$ 
 for radio-loud quasar activity}
\label{s: eff_lum}
In our model of quasar self-regulation, 
we rely upon interstellar cosmic rays generated in the radio 
core of quasar jets. As discussed in \S\ref{sec:jetmodel}, in the jet dissipation region a fraction $\epsilon_\ditto{th}\sim0.1$ of the jet's bulk kinetic power is converted into random thermal form by internal shocks. The average comoving energy of shocked protons will be in the GeV range, since their mean Lorentz factor is $\gamma_p\simeq1+\epsilon_\ditto{th}$; this corresponds in the host-galaxy frame to a kinetic energy $\simeq\Gamma\,\gamma_p\,m_pc^2\sim10\unit{GeV}$. We now address the question of whether a sufficient number of these \emph{moderately-relativistic} cosmic rays can diffuse out of the jet into the host galaxy, thus providing the coupling between jet power and interstellar gas required for our feedback scenario.

We find that most of the randomized protons produced by internal shocks are convected with the jet flow away from the dissipation region without suffering significant losses. Only a small fraction ($\sim t_\ditto{adv}/t_\ditto{diff,J}$) can escape from the jet before being advected into regions where magnetic irregularities in the flow are too weak to allow for significant particle diffusion. Here, $t_\ditto{adv}$ is the advection time through the dissipation region and $t_\ditto{diff,J}$ is the diffusion time across the jet, as measured in the jet comoving frame.\footnote{We point out that all timescales in this section are computed in the jet comoving frame.} It follows that the fraction of jet kinetic luminosity $L_\ditto{J}$ ending up as interstellar cosmic ray power will be
\be\label{eq:cr_eff}
\epsilon_\ditto{CR}\simeq\epsilon_\ditto{th}\frac{t_\ditto{adv}}{t_\ditto{diff,J}}~.
\ee
It is the purpose of this section to estimate the cosmic ray efficiency $\epsilon_\ditto{CR}$ and the corresponding interstellar cosmic ray power $L_\ditto{CR}=\epsilon_\ditto{CR}L_\ditto{J}$. Comparison of $L_\ditto{CR}$ with the Eddington limit in cosmic rays defined in eq.~(\ref{e: LeddCR}) will assess if enough cosmic rays are injected into the host galaxy to initiate a momentum-driven wind capable of removing the galactic gas, thus potentially self-regulating  the black hole -- galaxy co-evolution. 

\vspace{-0.1in}
\subsection{Important Timescales for $1-10\unit{GeV}$ Cosmic Rays in the Jet Dissipation Region}\label{sec:timescales}
Table \ref{tab1} summarizes the important comoving timescales for 1-10 GeV cosmic rays in the jet dissipation region; in boldface, the timescales most relevant for our model. As discussed below, we find that: \textit{i}) randomized relativistic protons are produced by internal shocks on a much shorter timescale than advection, which is itself the fastest loss process for 1-10 GeV ions; \textit{ii}) cosmic ray diffusion out of the jet is a slow process compared to advection, implying that only a small fraction of the shock-accelerated protons will contribute to the interstellar cosmic ray population.

\vspace{-0.1in}
\subsubsection{Cosmic Ray Production and Losses}\label{sec:loss}
Although eq.~(\ref{eq:tau_diss}) implies that the dissipation region 
is collisionless, the presence of even modest magnetic 
fields reasonably ensures that  the transition from the fast cold pre-shock 
flow to the relatively-slow hot post-shock medium occurs within a
few Larmor scales \citep[as observed for the Earth's bow shock by, e.g.,][]{tanaka_83,sckopke_90}. In the jet comoving frame, this implies that internal shocks heat and randomize the incoming cold protons on the short timescale
\begin{equation}\label{eq: t_heat}
t_\ditto{heat}\!\simeq\omega_\ditto{L}^{-1}\!\simeq3.9\times10^{-5}\,\gamma_p\frac{\Gamma_1^{3}\,\theta_\ditto{\min}}{\Lambda_{_{\mr{Edd}}}^{1/2}\,\epsilon_\ditto{B,\min}^{1/2}\,\epsilon_\ditto{th,\min}^{1/2}\epsilon_\ditto{rad,\min}} \,M_{\bullet,9}^{1/2}\unit{s}~,
\end{equation}
where $\omega_\ditto{L}=eB'/(\gamma_p m_p c)$ is the Larmor frequency of shock-heated protons with characteristic Lorentz factor $\gamma_p$; $B'$ is the comoving magnetic field strength in \eq{Bfield}. We remark that our cosmic ray-driven feedback model does not require highly-relativistic protons, possibly accelerated at internal shocks by some Fermi mechanism. Rather, the bulk of the self-regulation effect results from ``thermal'' protons with comoving energies in the GeV range, produced in the dissipation region on the timescale $t_\ditto{heat}$ computed above. For the sake of completeness, we also include in Table \ref{tab1} the characteristic timescale $t_\ditto{DSA}$ of diffusive shock acceleration.\footnote{Diffusive shock acceleration occurs on a timescale $t_\ditto{DSA}\sim t_\ditto{adv}/\tau_\ditto{J}$, where $\tau_\ditto{J}\gtrsim1$ is the cosmic ray optical depth  in the jet dissipation region (see \S \ref{sec:diff}). Therefore, although not required in our model, there would be enough time to accelerate a fraction of the shock-heated protons to suprathermal energies before the flow is advected away from the shock.}

Production of $\sim\rm{GeV}$ cosmic rays on the timescale $t_\ditto{heat}$ estimated above is much faster than advection through the dissipation region, which happens on a characteristic time
\begin{eqnarray}\label{eq:t_adv}
\!\!\!\!\!\!\!\!\!t_\ditto{adv}\!\!\simeq\!\frac{R_\ditto{diss}/\Gamma}{c}\!\sim\!\frac{R_\ditto{diss}\theta}{c}
\!\simeq\!4.9\times10^5\,\epsilon_\ditto{rad,\min}^{-1} \Gamma^2_1\theta_\min M_{\bullet,9}\unit{s}~,
\end{eqnarray}
where we have made use of the fact that in the comoving frame the dissipation region is roughly spherical ($R_\ditto{diss}/\Gamma\sim R_\ditto{diss}\theta$). This
 is of the same order as the adiabatic cooling time due to the jet lateral expansion:
\begin{eqnarray}\label{eq:t_exp}
t_\ditto{ad}\simeq\frac{1}{\beta_\ditto{exp}}\frac{R_\ditto{diss}\theta}{c}
\simeq4.9\times10^5\,\epsilon_\ditto{rad,\min}^{-1} \,\Gamma_1\,M_{\bullet,9}\unit{s}~,
\end{eqnarray}
where $\beta_\ditto{exp}\simeq\Gamma\theta$ is the jet expansion velocity in the comoving frame. 

As Table \ref{tab1} shows, advection and adiabatic cooling are the fastest loss processes for 1-10 GeV ions. The table also includes the characteristic timescales for the following processes: radiative cooling via synchrotron and inverse Compton emission, inelastic collisions with other protons ($p\,p$ collisions, with cross section $\sigma_{pp}\simeq3\times10^{-26}\unit{cm^2}$ at energies of a few GeV and inelasticity $k_{pp}\simeq1/2$) and inelastic collisions with background photons ($p\,\gamma$ collisions) -- resulting in photo-meson production and/or pair production via the Bethe-Heitler effect \citep[for an estimate of the corresponding cross section, see][]{sikora_87}. We also discuss the limiting  case in which the proton thermal energy stored in the dissipation region (see $U'_\ditto{th}$ in \eq{eps_p}) is efficiently transferred to the emitting electrons, which then cool via synchrotron and inverse Compton at the observed bolometric luminosity $L_\ditto{Bol}\sim10^{48}\unit{erg\,s^{-1}}$; the corresponding timescale is a conservative lower limit which holds regardless of uncertainties in the efficacy of energy exchange between protons and electrons.

\vspace{-0.22in}
\begin{table}[htbp]
\caption{Important Timescales for $1-10\unit{GeV}$ Protons in the Jet
  Dissipation Region}
\centering
\begin{tabular}{|l|c|c|}
\hline
\msc{PHYSICAL PROCESS} & \msc{TIMESCALE (s)} & \msc{SYMBOL}\\ [0.6ex]
\hline
\hline 
\msc{Shock Heating} & $\mathbf{3.9\times10^{-5}}\,\gamma_p$ & $t_\ditto{heat}$\\[0.5ex]
\hline
\msc{Fermi Acceleration} & $2.1\times10^{2}\,\gamma_p^{1/3}$ & $t_\ditto{DSA}^\uppo{(Kolm)}$\\[0.6ex]
&  $6.9\times10^{4}$ & $t_\ditto{DSA}^\uppo{(str)}$\\[0.6ex]
\hline
\hline
\msc{Advection} & $\mathbf{4.9\times10^{5}}$ & $t_\ditto{adv}$\\[0.5ex]
\hline
\msc{Adiabatic Cooling}& $\mathbf{4.9\times10^{5}}$ & $t_\ditto{ad}$\\ [0.5ex]
\hline
\msc{Synchrotron and IC Cooling}& $8.8\times10^{15}\,\gamma_p^{-1}$ & $$\\[0.5ex]
\hline
\msc{$p-p$ Inelastic Collisions}& $1.1\times10^{11}$ & $$\\[0.5ex]
\hline
\msc{$p-\gamma$ Inelastic Collisions}& $2.4\times10^{12}\,\gamma_p^{-1}$ & $$\\[0.5ex]
\hline
\msc{$p\rightarrow e$ Energy Transfer}& $1.7\times10^{6}$ & $$\\[0.5ex]
\hline
\hline
\msc{Diffusion} & $1.1\times10^{9}\,\gamma_p^{-1/3}$ & $t_\ditto{diff,J}^\uppo{(Kolm)}$\\[0.6ex]
&  $\mathbf{3.5\times10^{6}}$ & $t_\ditto{diff,J}^\uppo{(str)}$\\[0.6ex]
\hline
\end{tabular}
\footnotetext{\,When evaluating the above timescales,
we use $\Lambda_\ditto{Edd}=1$, $\Gamma=10$, $\theta=0.1$,
$\epsilon_\ditto{rad}=0.1$, $\epsilon_\ditto{th}=0.1$,
$\epsilon_{\ditto{B}}=0.1$ and $M_\bullet=10^9\,M_\sun$. We also assume that $L_\ditto{IC}/L_\ditto{S}=10$. In boldface, the
timescales most relevant for our model.}
\label{tab1}
\end{table}
\vspace{-0.15in}

\subsubsection{Cosmic Ray Diffusion out of the Jet}\label{sec:diff}
In the tangled and inhomogeneous fields of the dissipation region, a fraction of the shock-heated protons may efficiently scatter with resonant magnetic fluctuations and diffuse out of the jet before being advected away with the jet flow. The cosmic ray diffusion timescale across the jet  is given by
\be\label{eq:diff_time}
t_\ditto{diff,J}\simeq\tau_{\ditto{J}}\frac{R_\ditto{diss}\theta}{c}\sim\tau_{\ditto{J}}\,t_\ditto{adv}~,
\ee
where $\tau_\ditto{J}$ is the cosmic ray optical depth.\footnote{The hierarchy between the characteristic timescales for diffusion ($t_\ditto{diff,J}$), advection ($t_\ditto{adv}$) and diffusive shock acceleration ($t_\ditto{DSA}$) is such that $t_\ditto{diff,J}\sim\tau_\ditto{J}\,t_\ditto{adv}\sim\tau_\ditto{J}^2\,t_\ditto{DSA}$, and since $\tau_\ditto{J}\gtrsim1$ we have $t_\ditto{diff,J}\gtrsim t_\ditto{adv}\gtrsim t_\ditto{DSA}$.} The resonant magnetic fluctuations providing the cosmic ray scattering may be embedded in the jet flow with, e.g.,  a Kolmogorov spectrum, or generated in the dissipation region by the accelerated cosmic rays themselves via the so called ``streaming instability'' \citep[e.g.,][]{kulsrud_69,wentzel_74}. 

A concise review of both scattering processes and their ability to describe the cosmic ray distribution function in the Milky Way is given in  \S 3 of \citet{socrates_06}. 
We assume that the scattering mechanisms which account for the interstellar cosmic ray optical depth in the Galaxy operate in the core of quasar jets as well and we employ the same scalings as in \citet{socrates_06}. This may seem as a huge extrapolation, but in the absence of direct observational constraints on the cores of radio-loud quasars, we utilize this assumption for lack of a better choice. We now estimate the value of $\tau_\ditto{J}$ expected for the two scattering processes mentioned above.

The cosmic ray optical depth in the presence of resonant Kolmogorov turbulence at the Larmor scale  of shock-heated protons ($r_\ditto{L,J}=c/\omega_\ditto{L}$) may be written
\begin{equation}\label{eq:tau_kolm}
\tau_\ditto{J}^\uppo{(Kolm)}\simeq\left(\frac{\delta B_{\mr{0}}}{B'}\right)^2\left(\frac{R_\ditto{diss}\theta}{r_\ditto{L,J}}\right)^{1/3}\left(\frac{R_\ditto{diss}\theta}{\lambda_{\mr{0}}}\right)^{2/3}~,
\end{equation}
where $\lambda_{\mr{0}}$, $\delta B_{\mr{0}}$ and $B'$ are respectively the stirring scale of the magnetic turbulence, the fluctuation amplitude at the stirring scale and the jet magnetic field computed in \eq{Bfield}. Assuming $\delta B_{\mr{0}}\sim B'$ and $\lambda_{\mr{0}}\sim R_\ditto{diss}\theta$, \eq{tau_kolm} gives
\begin{equation}\label{eq:tau_kolm_j}
\tau_\ditto{J}^\uppo{(Kolm)}\!\simeq2.3\times10^3\,\gamma_{p}^{-1/3}\Lambda_{_{\mr{Edd}}}^{1/6}\epsilon_\ditto{B,\min}^{1/6}\epsilon_\ditto{th,\min}^{1/6}\Gamma^{-1/3}_1 M_{\bullet,9}^{1/6}~,
\end{equation}
and the corresponding diffusion time across the jet is
\begin{equation}
t_\ditto{diff,J}^\uppo{(Kolm)}\!\!\simeq\!\!1.1\times10^9\gamma_{{p}}^{-1/3}\Lambda_{_{\mr{Edd}}}^{1/6} \epsilon_\ditto{B,\min}^{1/6}\epsilon_\ditto{th,\min}^{1/6}\epsilon_\ditto{rad,\min}^{-1}\Gamma^{5/3}_1\theta_\ditto{\min} M_{\bullet,9}^{7/6}\unit{s}~.
\end{equation}

In the case that the cosmic ray opacity is provided by  magnetic fluctuations self-generated via the streaming instability, the optical depth may be written
\begin{equation}\label{eq:tau_str_j}
\tau_\ditto{J}^\uppo{(str)}\simeq\frac{c}{v_{_{A,\rm J}}}\simeq7.1\,\epsilon_\ditto{B,\min}^{-1/2}\epsilon_\ditto{th,\min}^{-1/2}~,
\end{equation}
where $v_{_{A,\rm J}}=B'/\sqrt{4\pi\rho'_\ditto{diss}}$ is the Alfv\'en velocity in the dissipation region. The diffusion time across the jet is then
\begin{equation}
t_\ditto{diff,J}^\uppo{(str)}\simeq3.5\times10^6\,\epsilon_\ditto{B,\min}^{-1/2}\,\epsilon_\ditto{th,\min}^{-1/2}\,\epsilon_\ditto{rad,\min}^{-1}\,\Gamma^2_{1}\,\theta_\ditto{\min}\,M_{\bullet,9}\unit{s}~.
\end{equation}

By modeling the spectral energy distribution of powerful blazars, \citet{celotti_ghisellini_07} infer that the $\sim0.5-10\unit{GeV}$ electrons responsible for the observed synchrotron and inverse Compton emission should be injected at internal shocks with a power-law energy spectrum $n(\gamma_e)\propto\gamma_e^{-\alpha}$ with $\alpha\simeq2.5$. If the mechanism responsible for the electron acceleration to suprathermal  energies depends only on particle rigidity, as is the case for a Fermi process, then a comparable spectral index should describe the accelerated protons. For first-order Fermi acceleration at mildly-relativistic internal shocks this implies, for $\sim\rm{GeV}$ protons, that $\tau_\ditto{J}\Delta E/E\sim\tau_\ditto{J}\Delta \Gamma/\Gamma\sim1$, since the inferred particle spectrum has nearly equal energy per logarithmic interval. Here,  $\Delta E/E$ is the average fractional energy gain per acceleration cycle. It follows that $\tau_\ditto{J}\sim\Gamma/\Delta\Gamma\gtrsim1$.

\subsection{Cosmic Ray Efficiency $\epsilon_\ditto{CR}$ and Luminosity $L_\ditto{CR}$}\label{sec:Lcr}
Armed with a better knowledge of the physical processes relevant for 1-10 GeV protons produced in the jet core, we can now compute the fraction $\epsilon_\ditto{CR}$ of jet power injected into the host galaxy as cosmic rays, and the corresponding interstellar cosmic ray luminosity $L_\ditto{CR}$. 

Since the fraction of shock-heated protons that can diffuse out of the jet before advection takes place is $\sim t_\ditto{adv}/t_\ditto{diff,J}\sim 1/\tau_{\ditto{J}}$, as shown in \eq{diff_time}, the cosmic ray efficiency in \eq{cr_eff} may be rewritten as $\epsilon_{\ditto{CR}}\simeq\epsilon_\ditto{th}/\tau_\ditto{J}$. If the cosmic ray scattering is due to resonant magnetic fluctuations with a Kolmogorov spectrum (cf. \eq{tau_kolm_j}), then
\be\label{eq:epscr_kolm}
\!\!\!\!\!\!\!\epsilon_\ditto{CR}^\uppo{(Kolm)}\!\!\!\!\simeq\!\frac{\epsilon_\ditto{th}}{\tau_\ditto{J}^\uppo{(Kolm)}}\simeq4.4\times10^{-5}\,\gamma_{p}^{1/3}\frac{\epsilon_\ditto{th,\min}^{5/6}\Gamma^{1/3}_1}{\Lambda_{_{\mr{Edd}}}^{1/6}\epsilon_\ditto{B,\min}^{1/6}} M_{\bullet,9}^{-1/6}~,
\ee
where $\gamma_p\sim1-10$ may be thought of as the characteristic comoving Lorentz factor of shock-accelerated protons. Instead, if the cosmic ray opacity is provided by magnetic waves self-generated via the  streaming instability (cf. \eq{tau_str_j}), the resulting cosmic ray efficiency is 
\be\label{eq:epscr_str}
\epsilon_\ditto{CR}^\uppo{(str)}\!\!\simeq\!\frac{\epsilon_\ditto{th}}{\tau_\ditto{J}^\uppo{(str)}}\simeq1.4\times10^{-2}\,\epsilon_\ditto{B,\min}^{1/2}\epsilon_\ditto{th,\min}^{3/2}~.
\ee
As eqs.~(\ref{eq:epscr_kolm}) and (\ref{eq:epscr_str}) suggest, the cosmic ray coupling efficiency $\epsilon_\ditto{CR}= L_\ditto{CR}/L_\ditto{J}$ is either weakly dependent (for Kolmogorov turbulence) or not dependent at all (for the streaming instability) on the black hole mass $M_\bullet$. In other words, our cosmic ray-driven feedback scenario is manifestly \emph{self-similar} or \emph{scale-independent} \textit{per unit of jet power} (or equivalently, per unit of black hole mass, for fixed radiative efficiency). 

The cosmic ray luminosity $L_\ditto{CR}=\epsilon_\ditto{CR}L_\ditto{J}$ injected into the host galaxy is then, for the two scattering mechanisms discussed above, 
\begin{eqnarray}\label{eq:l_cr_kolm}
\!\!\!\!\!\!\!\!L_\ditto{CR}^\uppo{(Kolm)}\!\!\!&\simeq&5.7\times10^{42}\gamma_p^{1/3}\frac{\Lambda_{_{\mr{Edd}}}^{5/6}\epsilon_\ditto{th,\min}^{5/6}\Gamma^{1/3}_1}{\epsilon_\ditto{B,\min}^{1/6}} M_{\bullet,9}^{5/6}\unit{erg\,s^{-1}}~,\\
\!\!\!\!\!\!\!\!L_\ditto{CR}^\uppo{(str)}&\simeq&1.8\times10^{45}\,\Lambda_{_{\mr{Edd}}}\epsilon_\ditto{B,\min}^{1/2}\epsilon_\ditto{th,\min}^{3/2}M_{\bullet,9}\unit{erg\,s^{-1}}~,\label{eq:l_cr_str}
\end{eqnarray}
where we have made use of $L_\ditto{J}=\Lambda_{\ditto{Edd}}L_\ditto{Edd,\bullet}$.

Comparison of eq.~(\ref{eq:l_cr_str}) with the Eddington limit in cosmic rays defined in eq.~(\ref{e: LeddCR}) (or equivalently, of eq.~(\ref{eq:epscr_str}) with eq.~(\ref{e: min_eff})) suggests that, if the resonant magnetic turbulence for cosmic ray scattering in the jet core is mainly generated by the streaming instability, there may be enough power in interstellar cosmic rays to launch a momentum-driven wind which could eject the galactic gas. Although the \textit{momentum} requirement $L_\ditto{CR}\gtrsim L_\ditto{Edd,CR}$ is a necessary condition for efficient self-regulation of the black hole -- galaxy system, it is not sufficient by itself; in fact, we must also require that the \emph{time-integrated} cosmic ray \emph{energy} injected into the host galaxy during this epoch of super-Eddington cosmic ray activity should be comparable to the binding energy of the gas. In \S\ref{sec:feedback} we address this important issue.

\subsection{Cosmic Ray Propagation within the Host Galaxy}\label{sec:propag}
In the frame of the host galaxy, the shock-accelerated protons which diffuse out of the jet are relativistically beamed  along the jet axis within an angle $\sim1/\Gamma$. Since the cosmic ray  optical depth for scattering off resonant magnetic fluctuations in the galaxy's interstellar medium is very large ($\tau_g\sim10^3$, as we show below), the cosmic ray momentum distribution quickly isotropizes so that their overall effect on the galactic gas resembles a spherically-symmetric pressure force in the direction opposite to the galaxy's gravitational center. As the interstellar cosmic rays diffuse towards larger scales, their energy does not appreciably change from the value $\simeq\Gamma\gamma_pm_pc^2\sim1-100\unit{GeV}$ of their birth, since the average fractional energy loss per scattering is only $\sim\tau_g^{-2}\ll1$ \citep[e.g.,][]{kulsrud_69,wentzel_74}. Instead, the fractional momentum change per scattering is $\sim\tau_g^{-1}$, which allows cosmic rays to be extremely effective in powering a momentum-driven outflow of interstellar gas. In this section, we study the cosmic ray propagation within the host galaxy, providing an estimate for the cosmic ray optical depth $\tau_g$ and diffusion time $t_\ditto{diff,\dg}$.

The characteristic galactic scale radius $R_g$ for an isothermal sphere is  
\begin{equation}
R_g\!\simeq\!\frac{G}{2}\frac{M_\star}{\sigma_\star^2}\!\simeq18.2\!\left(\frac{M_\star/M_\bullet}{10^{3}}\right)\left(\frac{M_{\bullet,9}}{0.76\, \sigma_{\star,300}^4}\right)\sigma^2_{\star,300}\unit{kpc}~,
\end{equation}
if the black hole -- galaxy system follows the $M_\bullet-M_\star$   and $M_\bullet-\sigma_\star$ relations. We take $R_g$ as the characteristic distance where cosmic rays produced in the jet core interact with the galactic gas.

If we pin the ratio between cosmic ray mean free path and Thomson mean free path at the galactic scale $R_g$ to the  value $\lambda_\ditto{CR}/\lambda_\ditto{T}\sim10^{-6}$ appropriate for the Milky Way, the cosmic ray optical depth up to $R_g$ is 
\be\label{eq:tau_g}
\!\!\!\!\!\tau_g\simeq \kappa_{es}\,\rho_gR_g\frac{\lambda_\ditto{T}}{\lambda_\ditto{CR}}\simeq1.5\times10^3f_{g,\min}\left(\frac{\lambda_\ditto{T}/\lambda_\ditto{CR}}{10^{6}}\right)~,
\ee
where $\rho_g$ is the gas mass density at $R_g$ for an isothermal sphere. 
The cosmic ray diffusion time up to $R_g$ is then
\be\label{eq:diff_gal}
\!\!\!\!\!\!\!\!\!\!t_\ditto{diff,\dg}\!\!\simeq\!\tau_g\frac{R_g}{c}\simeq8.9\times10^7f_{g,\min}\!\left(\frac{\lambda_\ditto{T}/\lambda_\ditto{CR}}{10^{6}}\right)\!\sigma^2_{\star,300}\unit{yr}~.
\ee
We expect that, during epochs of vigorous jet-powered cosmic ray activity, the galactic structure will be appreciably affected by the cosmic ray pressure on timescales comparable to the cosmic ray diffusion time computed in \eq{diff_gal}. It follows that, if the radio jet is powered by gas accretion onto the central black hole, then $t_\ditto{diff,\dg}$ might be a reasonable upper limit for the duration of the radio-loud phase. In \S \ref{sec:rp} we further comment on this and examine the implications of the scaling $t_\ditto{diff,\dg}\propto\sigma_\star^2$.

As discussed by \citet{socrates_06}, the cosmic ray pressure force may be reduced by $\sqrt{t_{pp,g}/t_\ditto{diff,\dg}}$ if the diffusion timescale $t_\ditto{diff,\dg}$ is significantly longer than the time $t_{pp,g}$ required to deplete the cosmic ray energy via inelastic scattering with background protons resulting in pion production. For a cross section $\sigma_{pp}\simeq3\times10^{-26}\unit{cm^2}$ at $\sim\rm{GeV}$ energies and a scattering inelasticity $\kappa_{pp}\simeq1/2$, we find 
\be
t_{pp,g}\simeq\frac{m_p}{\kappa_{pp}\sigma_{{pp}}c}\frac{1}{\rho_g}\simeq1.8\times10^8\,f_{g,\min}^{-1}\,\sigma^2_{\star,300}\unit{yr}~,
\ee
which is marginally longer than $t_\ditto{diff,\dg}$, so that we can ignore losses due to pion production.\footnote{We point out that cosmic ray losses due to $p\,p$ collisions are negligible all the way down to pc scales. If the Thomson optical depth in the BLR is $\tau_\ditto{T}\sim0.1$, the extreme assumption that the cosmic ray diffusion time in the BLR be comparable or longer than the proton-proton collision time implies a CR optical depth at pc scales $\tau_\ditto{CR}\gtrsim m_p\kappa_{es}/(\kappa_{pp}\sigma_{pp})\,\tau_\ditto{T}^{-1}\simeq450$. In the BLR, the gravitational force is still primarily provided by the black hole, and the correspoding Eddington luminosity in CRs will be $L_\ditto{Edd,CR}=L_\ditto{Edd,\bullet}\tau_\ditto{T}/\tau_\ditto{CR}\simeq2.0\times10^{43}M_{\bullet,9}\unit{erg\unit{s^{-1}}}$. This is smaller than the CR luminosity in \eq{l_cr_str}. Therefore,  in this case, the CR flux would disrupt the entire BLR. Since the BLR \textit{exists}, we can comfortably reject the extreme assumption that CRs are significantly destroyed at pc scales.} 

Finally, we show that cosmic ray optical depths comparable to the value in \eq{tau_g} may be derived by making the extreme assumption that  the magnetic energy density in the galaxy is roughly in equipartition with the gas random kinetic energy density, i.e. $B^2/8\pi\sim3/2\,\rho_{g}\sigma_\star^2$ at the characteristic radius $R_g$. If the cosmic ray scattering is provided by background magnetic turbulence with a Kolmogorov spectrum, the optical depth up to $R_g$ is
\begin{equation}
\tau_g^\uppo{(Kolm)}\!\!\!\!\!\!\simeq\left(\frac{\delta B_{\mr{0}}}{B}\right)^2\!\left(\frac{R_g}{r_\ditto{L,\dg}}\right)^{1/3}\!\!\left(\frac{R_g}{\lambda_{\mr{0}}}\right)^{2/3}\!\!\!\!\!\!\simeq4.4\times10^3f_{g,\min}^{1/6}\,\sigma^{2/3}_{\star,300}~,
\end{equation}
where $\lambda_0\sim R_g$ is the stirring scale and $\delta B_0\sim B$ the fluctuation amplitude at the stirring scale for the interstellar magnetic turbulence, and $r_\ditto{L,\dg}$ is the Larmor radius in the galactic magnetic field $B$ for a $\sim10\unit{GeV}$ proton. Instead, if the cosmic ray opacity results from Alfv\'en waves self-generated via the streaming instability, the cosmic ray optical depth will be 
\begin{equation}
\tau_g^\uppo{(str)}\simeq\frac{c}{v_{_{A,\dg}}}\simeq5.8\times10^2\sigma^{-1}_{\star,300}~,
\end{equation}
where $v_{_{A,\dg}}=B/\sqrt{4\pi\rho_g}$ is the Alfv\'en velocity in the galaxy. Note that both scattering mechanisms yield an optical depth comparable to the result in \eq{tau_g}.

\section{Cosmic Ray Feedback Resulting from Episodic Radio-Loud Activity: Origin of the $M_{\bullet}-M_{\star}$ Relation}
\label{sec:feedback}
As discussed in \S\ref{sec:Lcr}, during epochs of powerful jet activity the cosmic ray luminosity $L_\ditto{CR}$ injected into the interstellar medium of the host galaxy may exceed the galaxy's Eddington limit in cosmic rays $L_\ditto{Edd,CR}$ (compare eqs.~(\ref{eq:l_cr_str}) and (\ref{e: LeddCR})). Due to the outward cosmic ray pressure force, hydrostatic balance is lost and a cosmic ray momentum-driven wind develops, removing the fuel for further  star formation and black hole accretion. However, the bulge and black hole growth is completely choked off only if the radio-loud phase lasts long enough such that the total energy output in cosmic rays, which would eventually couple to the interstellar medium, is comparable to the binding energy of the galactic gaseous component. In other words, the \textit{momentum} requirement $L_\ditto{CR}\gtrsim L_\ditto{Edd,CR}$ is a necessary prerequisite, but efficient self-regulation of the black hole -- galaxy system is achieved only when the \textit{energy} requirement discussed above is fulfilled as well.

A generic feedback mechanism is capable of self-regulating the combined black hole -- galaxy growth  only if the total energy injected into the interstellar medium $\Delta E_{\ditto{inj}}$  is comparable to the gravitational energy of the gas: 
\be\label{eq:en_feed}
\Delta E_{\ditto{inj}}\sim E_g\simeq f_gM_\star\sigma_\star^2~,
\ee
where we have adopted an isothermal sphere. In what follows, we assume that the gas fraction $f_g$ 
is independent of black hole mass and stellar velocity dispersion. We now contrast our jet-powered cosmic ray feedback model with self-regulation mechanisms that act during the optically-luminous ``quasar phase.''

\subsection{Black Hole Self-Regulation during the Luminous Quasar Phase}\label{sec:qp}
\begin{figure}[htp]
\centering
\includegraphics[width=0.5\textwidth]{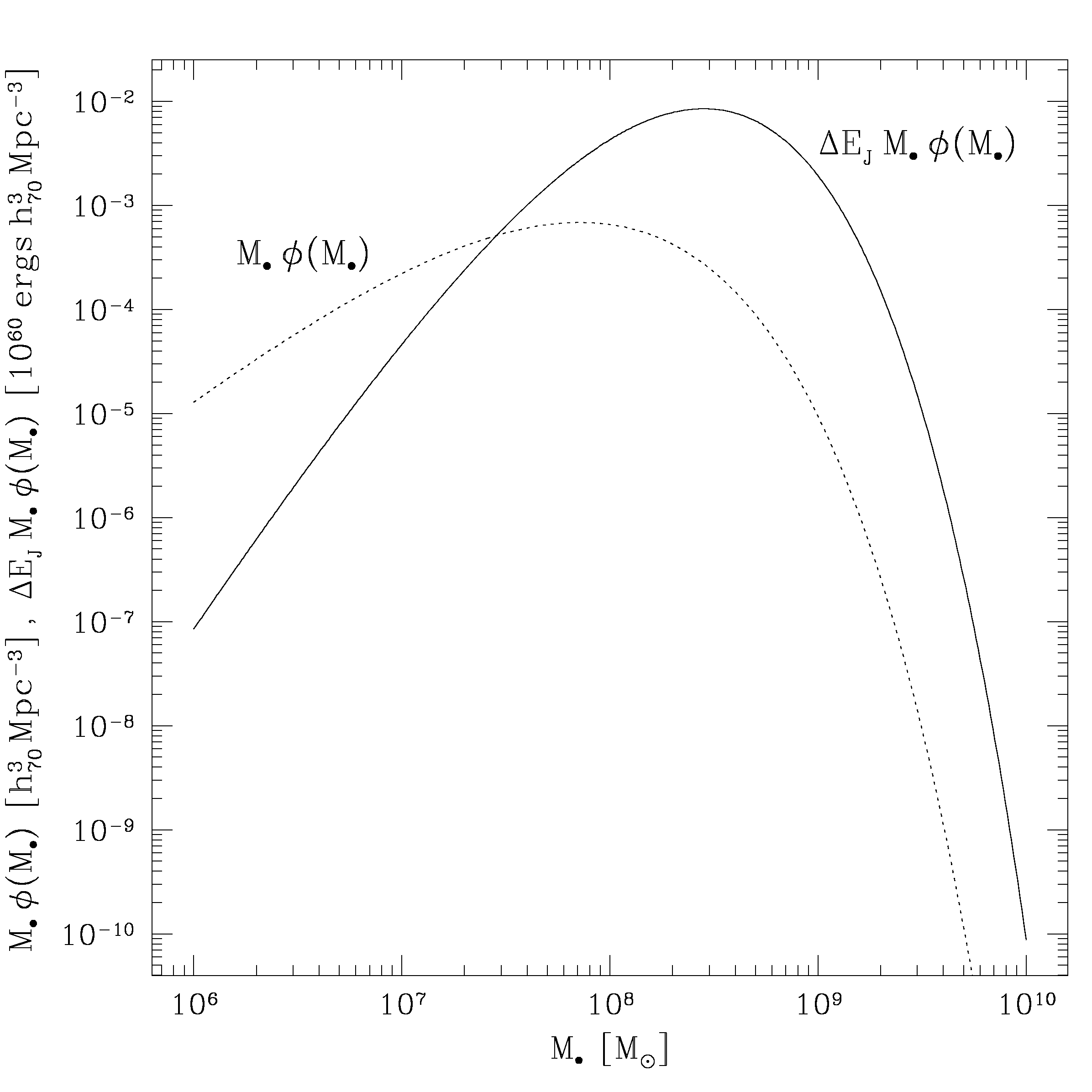}
 \caption{Black-hole mass function ($M_\bullet\,\phi(M_\bullet)$; dotted line) and kinetic energy output of radio jets integrated over cosmic time  ($\Delta E_{\ditto{J}}M_\bullet\,\phi(M_\bullet)$; solid line), versus black hole mass $M_\bullet$. The function $\phi(M_\bullet)$, namely the number density of black holes with mass in the interval $[M_\bullet,M_\bullet+\ud M_\bullet]$, is derived from the  stellar velocity dispersion function of early-type galaxies by \citet{bernardi_03} via the $M_\bullet-\sigma_\star$ relation.  The jet time-integrated kinetic  output $\Delta E_{\ditto{J}}$ is supposed to satisfy the energy balance $\epsilon_{\ditto{RP}}\Delta E_{\ditto{J}}\sim f_g M_\star\sigma_\star^2$ required for black hole self-regulation during radio-loud phases, as in \eq{balance_rp}. We have assumed a coupling efficiency  $\epsilon_{\ditto{RP}}=10^{-3}$, as found for our cosmic ray-driven feedback model (somewhat an intermediate value between \eq{epscr_kolm} and  \eq{epscr_str}) and a gas fraction $f_g=0.1$ for the host galaxy; the black hole -- galaxy system lies on the $M_\bullet-M_\star$ and $M_\bullet-\sigma_\star$ relations. We set $h_{70}$ to be the Hubble constant $H_0$ in units of $70\unit{km\,s^{-1}\,Mpc^{-1}}$.}
 \label{fig:ljet}
 \end{figure}

The \citet{soltan_82} argument, along with the work of \citet{yu_tremaine_02}, indicates that supermassive black holes at the center of galaxies build up their mass primarily by an act of radiatively-efficient accretion, in a relatively short-lived
high-luminosity ``quasar phase''.  During this epoch, the accretion luminosity approaches the black hole Thomson Eddington limit $L_{\ditto{Edd,\bullet}}$.\footnote{Caution must be taken when applying the \citet{soltan_82} argument to black hole 
demographics, encapsulated by the \citet{magorrian_tremaine_98}
relation.  The luminosity and mass function of supermassive black holes 
peak close to a black hole mass $M_\bullet\sim 10^8\,M_{\odot}$ (see dotted line in \fig{ljet}).
Relatively small black holes, like the one at the center of the Milky Way,
are both too rare and faint, such that a Soltan-like argument cannot be employed in order to determine whether 
or not their mass was built up by radiatively-efficient Eddington-limited
accretion.  However, it seems likely that 
relatively small black holes with masses as low as
$M_\bullet\simeq 10^6-10^7\,
M_{\odot}$ can radiate close to the Eddington limit, which therefore
suggests that the build-up of black hole mass in relatively 
small systems takes place 
during a bright short-lived ``quasar phase'' as well.}

If energy release during the optically-bright quasar phase is responsible for black hole self-regulation, the total energy $\Delta E_{\ditto{QP}}$ absorbed by the galactic gas during the quasar lifetime $\Delta t_\ditto{QP}$ should satisfy $\Delta E_{\ditto{QP}}\sim E_g$, as prescribed by \eq{en_feed}. If only a fraction $\epsilon_\ditto{QP}$ of the accretion luminosity $L_\ditto{acc}$ can couple to the interstellar medium of the host galaxy, we require
\be\label{eq:balance_qp}
\Delta E_{\ditto{QP}}\simeq\epsilon_\ditto{QP}\,\Delta t_\ditto{QP}\,L_\ditto{acc}\sim E_g~.
\ee
Assuming $L_\ditto{acc}\sim L_\ditto{Edd,\bullet}$, this can be rewritten as
\be\label{eq:magg1}
\frac{M_\bullet}{M_{\star}}\sim \left(\frac{\sigma_{_{\rm T}}}{
4\pi\,G\,m_p\,c}\right)\frac{f_g\,\sigma_\star^2}{
\epsilon_{_{\rm QP}}\,\Delta t_{_{\rm QP}}}~.
\ee
The quasar lifetime $\Delta t_\ditto{QS}$, set by the \citet{soltan_82} argument, is roughly comparable to the Salpeter time of the black hole:
\be
\!\!\!\!\!\!\!\!t_\ditto{Salp} \!= \! \frac{\Delta E_\bullet}{L_{_{\rm Edd,\bullet}}}
\!\simeq \!\left(\frac{\sigma_{_{\rm T}}\,c}{
4\pi\,G\,m_p}\right)\!\epsilon_{_{\rm rad}}\!\simeq\!4.5\times10^7\epsilon_\ditto{rad,\min}\unit{yr}~,
\label{e: Salpeter}
\ee
where $\Delta E_\bullet\simeq\epsilon_\ditto{rad}M_\bullet c^2$ is the total radiative energy output of the accretion process. Since $\epsilon_\ditto{rad}$ depends only on black hole spin, the quasar lifetime $\Delta t_\ditto{QS}\sim t_\ditto{Salp}$ should be independent of black hole mass. It follows from \eq{magg1} that, in order for a system to lie on the $M_\bullet-M_\star$ relation, the coupling efficiency $\epsilon_\ditto{QP}$ in the luminous quasar phase must depend on the stellar velocity dispersion $\sigma_\star$ such that
\be\label{eq:eff_qp}
\epsilon_\ditto{QP}\propto\sigma_\star^2~.
\ee
This implies that, if the quasar phase were to self-regulate the black hole and galaxy evolution, the feedback mechanism at work would \textit{not} be a \textit{universal}, \textit{scale-free} or \textit{self-similar}  process. We stress again that this follows from $\Delta t_\ditto{QP}\sim t_\ditto{Salp}$, or equivalently from the fact that the total quasar energy output $\Delta t_\ditto{QP}L_\ditto{acc}$ is observationally pinned to be close to the 
maximum amount of energy released in the formation of the black hole, given by $\Delta E_\bullet$.\footnote{The scaling in \eq{eff_qp} seems to be in contradiction with the simulations by \citet{dimatteo_05}, which reproduce the $M_\bullet-\sigma_\star$ relation assuming that a constant fraction of black hole accretion energy is deposited in the galactic gas \citep{springel_05}. However, we remark that $\epsilon_\ditto{QP}$ in \eq{eff_qp} is the fraction of  accretion energy \emph{available to unbind the galactic gas}, and not just the fraction of energy \emph{deposited in the gas}. If the simulations by Di Matteo et al. (2005) are taken at face value, then there is a hidden parameter that limits the fraction of deposited energy which is available to unbind the galactic gas. The hidden parameter should scale as $\propto\sigma_\star^2$.}

There is no apparent reason why the coupling efficiency $\epsilon_\ditto{QP}$
may not depend on the stellar velocity dispersion $\sigma_\star$ or the black hole mass $M_\bullet$. However, it should depend 
upon $\sigma_{\star}$ and $M_\bullet$ in such a way that makes
the ratio $M_\bullet/M_{\star}$ a constant across nearly four decades in black hole mass.  This is, of course, not impossible, but it would have to involve a cosmic 
conspiracy with respect to the gas dynamics of black hole 
self-regulation. We now show how our jet-powered cosmic ray feedback scenario provides a satisfactory solution for this apparent contradiction.

\subsection{Black Hole Self-Regulation During the Explosive 
Radio-Loud Phase}\label{sec:rp}

For a generic feedback mechanism operating during epochs of AGN radio-loud activity, the total energy $\Delta E_\ditto{RP}$ injected into the interstellar medium is
\be\label{eq:deltaEJ}
\Delta E_\ditto{RP}\simeq\epsilon_\ditto{RP}\Delta E_\ditto{J}\simeq\epsilon_\ditto{RP}\Delta t_{\ditto{RP}}L_\ditto{J}~,
\ee
where $\Delta E_\ditto{J}\simeq\Delta t_{\ditto{RP}} L_\ditto{J}$ is the time-integrated kinetic output of the radio jet, $\Delta t_{\ditto{RP}}$ is the duration of the radio-loud phase and $\epsilon_\ditto{RP}$ is the efficiency for coupling the jet power $L_\ditto{J}$ to the galactic gas. The energy balance $\Delta E_{\ditto{RP}}\sim E_g$ in \eq{en_feed}  requires that
\be\label{eq:balance_rp}
\Delta E_{\ditto{RP}}\simeq\epsilon_\ditto{RP}\,\Delta t_\ditto{RP}\,\Lambda_\ditto{Edd}L_{\ditto{Edd,\bullet}}\sim E_g~,
\ee
where $\Lambda_\ditto{Edd}=L_\ditto{J}/L_{\ditto{Edd,\bullet}}$. This can be rewritten as
\be\label{eq:magg2}
\frac{M_\bullet}{M_{\star}}\sim \left(\frac{\sigma_{_{\rm T}}}{
4\pi\,G\,m_p\,c}\right)\frac{f_g\,\sigma_\star^2}{
\epsilon_{_{\rm RP}}\,\Delta t_{_{\rm RP}}\,\Lambda_\ditto{Edd}}~,
\ee
from which it is apparent that the product $\epsilon_\ditto{RP}\,\Delta t_\ditto{RP}\,\Lambda_\ditto{Edd}$ for radio-loud phases replaces $\epsilon_\ditto{QP}\,\Delta t_\ditto{QP}$ in quasar epochs (cf. \eq{magg1}). For a black hole -- galaxy system that lies on the $M_\bullet-M_\star$ relation, this implies
\be\label{eq:radio_prop}
\epsilon_\ditto{RP}\Delta t_\ditto{RP}\,\Lambda_\ditto{Edd}\propto\sigma_\star^2~.
\ee
As opposite to the quasar feedback scenario discussed in \S\ref{sec:qp}, \eq{radio_prop} does not immediately constrain the coupling efficiency $\epsilon_\ditto{RP}$ to depend upon the stellar velocity dispersion. The possibility that the feedback efficiency 
$\epsilon_{_{\rm RP}}$ is independent of $M_\bullet$ and $\sigma_\star$ would then require that $\Delta t_{_{\rm RP}}\,\Lambda_{_{\rm Edd}}$
increases with the stellar velocity dispersion, or that the time-integrated jet kinetic output per unit of black hole mass should scale as $\propto\sigma_\star^2$. 

If black hole self-regulation in the radio-loud phase is mediated by cosmic rays produced in the jet core, we have shown that the coupling efficiency $\epsilon_\ditto{CR}$  is indeed independent (for cosmic rays scattering with Alfv\'en waves self-generated  via the streaming instability, see \eq{epscr_str}) or weakly dependent (for resonant diffusion by magnetic turbulence with a Kolmogorov  spectrum, see \eq{epscr_kolm}) on the black hole mass.\footnote{We have defined $\epsilon_\ditto{CR}$ as the fraction of jet power injected into the host galaxy as cosmic rays. However, due to the large cosmic ray optical depth within the host galaxy (see \S\ref{sec:propag}), the whole interstellar cosmic ray power will  eventually be transferred to the galactic gas, so that $\epsilon_\ditto{CR}$ is also a good proxy for the feedback coupling efficiency.} In our cosmic ray-driven feedback model we should then expect that $\Delta t_{_{\rm RP}}\,\Lambda_{_{\rm Edd}}\propto\sigma_\star^2$  or, if $\Lambda_\ditto{Edd}$ is a constant with black hole mass, that the duration of the radio-loud phase should scale as $\Delta t_\ditto{RP}\propto\sigma_\star^2$. 

As discussed in \S \ref{sec:propag}, a reasonable upper limit for $\Delta t_\ditto{RP}$ may be given by the cosmic ray diffusion timescale $t_\ditto{diff,\dg}$ within the host galaxy. In fact, if the jet is powered by black hole accretion, AGN radio activity would be terminated when the interstellar cosmic rays have diffused up to the scales where most of the gas resides, and started to push it outward, thus preventing further accretion. Interestingly, if  $f_g$ and $\lambda_\ditto{CR}/\lambda_\ditto{T}$ do not depend on black hole mass, \eq{diff_gal} shows that $t_\ditto{diff,\dg}\propto\sigma_\star^2$, the same scaling that $\Delta t_\ditto{RP}$ should have in order to satisfy \eq{radio_prop} with a scale-independent coupling efficiency. 

Note that for $\Delta t_\ditto{RP}\sim t_\ditto{diff,\dg}$, \eq{diff_gal} suggests that the duration of the radio-loud phase for the most massive galaxies may be comparable to the Salpeter time in eq.~(\ref{e: Salpeter}), whereas in relatively small galaxies it should last less than the quasar phase. In other words, for low-mass galaxies the black hole self-regulation would be confined to an episodic radio-loud epoch much shorter than the time required to accrue the black hole mass.

\section{Discussion and Comparison with Observations}\label{s: test}
We propose a self-regulation mechanism for supermassive black holes and their host bulges that operates in actively-accreting systems with powerful radio-loud activity. In the core of relativistic radio jets, internal shocks resulting from the dissipative interaction of shells of matter, intermittently ejected from the central engine, convert a fraction of the ordered kinetic energy of the jet flow into thermal form. If a sufficient number of the protons (or ``cosmic rays'') heated and randomized at internal shocks  escape the radio core into the interstellar medium of the host galaxy, they may profoundly affect the evolution of the galaxy and its black hole. To quantify their effect on the hydrostatic balance of the galactic gas we define an Eddington limit in cosmic rays for the host galaxy. For a phenomenologically-motivated jet model, we show that during powerful radio-loud phases the power in  1-100 GeV interstellar cosmic rays is large enough to break the cosmic ray Eddington limit for the host galaxy, so that \emph{momentum} balance of the galactic gaseous component is lost and a cosmic ray-driven wind develops, that removes the galactic gas.  If this super-Eddington cosmic ray activity lasts long enough, the time-integrated cosmic ray \textit{energy} input into the interstellar medium may exceed the binding energy of the gas, and the whole galaxy's gaseous phase will become unbound.  In doing so, any fuel for further black hole growth and star formation will be removed, thus affecting the combined evolution of the black hole and its stellar bulge, as implied by the $M_\bullet-M_\star$ and $M_\bullet-\sigma_\star$ relations. 

\citet{morganti_05} report the detection of \textit{fast} ($\sim1000\unit{km\,s^{-1}}$) \textit{large-scale} ($\sim1-10\unit{kpc}$) massive outflows of \textit{neutral} hydrogen in several powerful radio galaxies. They claim that the outflows are spatially associated with radio knots extended  along the jet, but this may just result from the fact that H~I absorption can be traced only in the presence of a bright background radio continuum. Fast large-scale outflows of low-ionization species with nearly \textit{spherical} morphology have been detected in the powerful radio source MRC~1138-262 by \citet{nesvadba_06}. The fact that the outflow is ``cold'' (atomic or weakly ionized) and ``fast'' (super-virial) implies that it is the result of catastrophic momentum, rather than energy, exchange with the interstellar medium. In other words, it is possible that some sort of Eddington limit is being broken. Among the momentum-driven feedback mechanisms, \citet{nesvadba_06} conclude that neither radiation-powered quasar winds \citep[e.g.,][]{fabian_99, king_03, murray_05} nor direct coupling between the jet and the interstellar medium in the ``dentist drill'' model \citep{begelman_cioffi_89} seem sufficient to explain the large-scale gas kinematics observed in MRC~1138-262. Direct coupling between the jet and the galactic gas should result  in the outflow being confined within a narrow cone around the jet axis, contrary to observations. Regarding the effect of radiation pressure from the quasar photon output on interstellar dust grains \citep[e.g.,][]{murray_05}, it is hard to understand how this could remove a significant fraction of the galactic gas, since the UV photons will be degraded to IR frequencies after a few scatterings, within a parsec from the black hole. Also, the optical depth for the resulting far IR photons will not exceed unity beyond a few tens of parsecs. 

The jet-powered cosmic ray feedback scenario presented here does not suffer any of these deficiencies. The cosmic ray luminosity injected into the host galaxy is smaller than the black hole photon power by roughly two orders of magnitude, but cosmic rays exchange momentum with the galactic gas $\sim\tau_g/\tau_\ditto{UV}\sim10^3$ more efficiently. Here, $\tau_{\ditto{UV}}\sim1$ is the optical depth for UV photons on dust grains. Moreover, since the  cosmic ray optical depth in the galaxy's interstellar medium is very large ($\tau_g\sim10^3$), any memory of the cosmic ray momentum distribution at injection is quickly lost and their effect on the galactic gas should resemble a spherically-symmetric outward pressure force, so that the resulting gas outflow is nearly spherical. Also, the fraction of cosmic ray energy lost at each interaction with the interstellar gas is minimal ($\sim\tau_g^{-2}$), meaning that they can propagate up to the large scales where most of the gas resides without suffering significant losses.

An interesting result of our study is that, per unit of black hole energy release, the cosmic ray feedback efficiency $\epsilon_\ditto{CR}$, which gives the fraction of jet power available to unbind the galactic gas, is roughly a constant with black hole mass. 
Instead, for self-regulation models relying on the  black hole photon output during radiatively-efficient quasar phases \citep[e.g.,][]{silk_rees_98, fabian_99, ciotti_ostriker_01, king_03,dimatteo_05,murray_05,hopkins_06}, the feedback efficiency is constrained to scale as $\sim E_g/\Delta E_\bullet\propto\sigma_\star^2$ in order for the system to lie on the $M_\bullet-M_\star$ relation. Here, $E_g\simeq f_g M_\star\sigma_\star^2$ is the binding energy of the galactic gas and $\Delta E_\bullet$ is the time-integrated energy output resulting from black hole accretion. It would be a surprising coincidence if an intrinsically \textit{scale-dependent} self-regulation mechanism were to result in the \textit{scale-free} $M_\bullet-M_{\star}$ relation, which holds for nearly four decades in mass.

The explosive radio-loud phase does not suffer from such severe constraints. Rather, since the coupling efficiency $\epsilon_\ditto{CR}$ for our cosmic ray-driven feedback scenario is scale-independent, the energy balance $\epsilon_{\ditto{CR}}\Delta E_{\ditto{J}}\sim E_g$ required for black hole self-regulation implies that the time-integrated jet kinetic output should scale as $\Delta E_{\ditto{J}}\propto M_\star \sigma_\star^2$ for a constant gas fraction $f_g$. By inferring the black hole mass function $M_\bullet\,\phi(M_\bullet)$ (dotted line in \fig{ljet}) from the velocity dispersion function of early-type galaxies by \citet{bernardi_03} via the $M_\bullet-\sigma_\star$ relation, we can predict the dependence on black hole mass of the kinetic energy output of radio jets integrated over cosmic time ($\Delta E_\ditto{J}M_\bullet\,\phi(M_\bullet)$; solid line in \fig{ljet}), for systems lying  on the $M_\bullet-M_\star$ and $M_\bullet-\sigma_\star$ relations. We find that its slope at the low-mass end is $\sim2.5$ and it peaks at $M_\bullet\simeq3\times10^8\,M_\sun$. With more reliable measurements of jet kinetic power and black hole mass, this could provide a stringent observational test for our proposed self-regulation scenario. For example, finding a slope close to 2.5 at the low-mass end would imply that the ratio between the time-integrated jet kinetic output and the binding energy of the galactic gas should be independent of black hole mass, thus strongly implicating that black hole self-regulation occurs in the radio-loud phase, irrespective of the actual coupling mechanism itself.

For a generic feedback mechanism acting during radio-loud epochs, the scaling $\Delta E_{\ditto{J}}\propto M_\star \sigma_\star^2$ can be recast as $\Delta t_{\ditto{RP}}\Lambda_\ditto{Edd}\propto \sigma_\star^2$, for a black hole -- galaxy system which follows the $M_\bullet-M_\star$ relation. Here, $\Lambda_\ditto{Edd}$ is the ratio of the jet kinetic power to the black hole Thomson Eddington limit  and $\Delta t_{\ditto{RP}}$ is the duration of the radio-loud phase. This implies that, for systems where a significant amount of mass is being built up, radio-loud signatures will most likely be observed in galaxies with large black holes and bulges, in line with the basic phenomenology of radio-loud AGN and quasars \citep[e.g.,][]{laor_00,best_05}. The $M_\bullet-M_{\star}$ relation perhaps is then connected to the fact that radio-loud actively-accreting objects are relatively absent in relatively small systems \citep{ho_06}: black hole  self-regulation takes place during the radio-loud phase 
and, due to the abundance of total energy available, this
phase is short-lived in 
relatively small systems because the amount of energy 
required to unbind the galaxy is relatively low.  

\acknowledgments{We thank Annalisa Celotti, Peter Goldreich, Jenny Greene, Jerry Ostriker and Anatoly Spitkovsky for helpful discussions.   
AS acknowledges support from a Lyman Spitzer Jr. Fellowship given by the Department of Astrophysical Sciences at  Princeton University and a Friends of the Institute Fellowship at the Institute for Advanced Study in Princeton.}

\bibliography{ms}

\end{document}